\documentstyle[preprint,aps]{revtex}

\tightenlines
\newtheorem{theorem}{Theorem}
\newtheorem{acknowledgement}[theorem]{Acknowledgement}

\begin{document}
\title{Energy Eigenvalues of Kemmer Equation for a Homogeneous Magnetic Field}
\author{A. Havare$^{\ast }$, K. Sogut$^{\dagger }$}
\address{* Mersin University Department of Physics, Mersin, Turkey\\
$\dagger $Cukurova University Department of Physics, Adana, Turkey}
\maketitle

\begin{abstract}
This article illustrates a completely algebraic method to obtain the energy
levels of a massive spin-1 particle moving in a constant magnetic field. In
the process to obtain the energy levels the wave function was written by
harmonic oscillator solutions.
\end{abstract}

\section{INTRODUCTION}

The relativistic wave equation of massive spin-1 particles was at first
derived by Kemmer in 1939\cite{1}. Kemmer theory gives an entirely different
aspect of the mesons. Kemmer equation is a Dirac type equation but involves
matrices obeying different scheme of commutation rules. These rules were
first given by Duffin \cite{2}. The theory can be developed in strikingly
close correspondence to Dirac's electron theory; practically all the
definitions of physical quantities like spin, magnetic moment etc. have
their exact counterpart.

The dynamics of the charged spin-1 particles interacting with an external
field has been investigated for many situations. There are various
techniques of description of spin-1 particles moving in an external field 
\cite{3}-\cite{6}. In addition to the complexity of these techniques, the
results obtained from them are quite different. It has been showed that all
these techniques are equal when there is no anomalous magnetic moment
coupling.

The generally accepted method of solving the wave equation of spinning
particles is to use the differential equation solving technique but this
method of approach becomes complicated when higher spinning particles are
handled. This article illustrates a completely algebraic method for finding
the energy levels of spin-1 particles moving in a constant magnetic field.
It does not require an explicit solution of the equation. Spin-1 particle
will be considered as a two-fermion system of equal masses. In the process
to obtain the energy levels of massive spin-1 particles the harmonic
oscillator solutions that are given as Hermite polynomials are used. To
introduce the method we will first apply\ it to find the energy spectrum of
spin-1/2 particles. In the conclisions part we compared our results with the
previous results and saw that they were in agreement.

The vector potential we used is 
\begin{equation}
\overrightarrow{A}=xB\widehat{j}.  \label{1}
\end{equation}
where B is constant.

\bigskip

\section{\protect\bigskip ENERGY SPECTRUM OF A SPIN-1/2 PARTICLE}

Covariant form of the Dirac equation for a spin-1/2 particle moving in a
constant magnetic field is 
\begin{equation}
\left( \gamma ^{\mu }\pi _{\mu }-m\right) \Psi _{D}\left( x\right) =0
\label{2}
\end{equation}
where $\gamma ^{\mu }$ are $4\times 4$ Dirac matrices; $\pi _{\mu }=\left(
P_{0}-eA_{0},\overrightarrow{P}-e\overrightarrow{A}\right) $ is
electrodynamical four-momentum vector; $m$ is the mass of spin-1/2 particle, 
$\Psi \left( x\right) $ is the four-component spinor; $A_{\mu }=\left( \Phi ,%
\overrightarrow{A}\right) $ is the four-vector electromagnetic potential and 
$e$ is the charge of the electron. In the writing of the equation the
Heaviside units $c=\hbar =1$ were used. The gamma matrices are given in the
form 
\begin{eqnarray}
\gamma ^{0} &=&\beta =\left( 
\begin{array}{cc}
\text{I} & 0 \\ 
0 & \text{I}
\end{array}
\right) \text{ \ , \ }\overrightarrow{\gamma }=\beta \overrightarrow{\alpha }
\label{3} \\
\overrightarrow{\alpha } &=&\gamma ^{0}\overrightarrow{\gamma }=\left( 
\begin{array}{cc}
0 & \overrightarrow{\sigma } \\ 
\overrightarrow{\sigma } & 0
\end{array}
\right) \text{ \ .}  \label{4}
\end{eqnarray}
Eq.$\left( 2\right) $ can be written as 
\begin{equation}
\left( \overrightarrow{\alpha }\cdot \overrightarrow{\pi }+\beta m+e\Phi
\right) \Psi \left( x\right) =\pi _{0}\Psi \left( x\right)  \label{5}
\end{equation}
For steady states if we choose the wave function in the form 
\begin{equation}
\Psi \left( x_{0}=t,\overrightarrow{x}\right) =e^{-iEt}\left( 
\begin{array}{c}
\varphi \left( \overrightarrow{x}\right) \\ 
\chi \left( \overrightarrow{x}\right)
\end{array}
\right)  \label{6}
\end{equation}
and use $\overrightarrow{\alpha }$ matrices, Eq.$\left( 5\right) $ takes the
following form 
\begin{equation}
\left( 
\begin{array}{cc}
m+e\Phi & \overrightarrow{\sigma }\cdot \overrightarrow{\pi } \\ 
\overrightarrow{\sigma }\cdot \overrightarrow{\pi } & -m+e\Phi
\end{array}
\right) \left( 
\begin{array}{c}
\varphi \left( \overrightarrow{x}\right) \\ 
\chi \left( \overrightarrow{x}\right)
\end{array}
\right) =E\left( 
\begin{array}{c}
\varphi \left( \overrightarrow{x}\right) \\ 
\chi \left( \overrightarrow{x}\right)
\end{array}
\right) \text{ \ .}  \label{7}
\end{equation}
Since the particle is moving in a magnetic field, $A_{0}=\Phi =0$. By
defining $\varphi $ and $\chi $ as 
\begin{equation}
\varphi \left( \overrightarrow{x}\right) =\left( 
\begin{array}{c}
\varphi _{+} \\ 
\varphi _{-}
\end{array}
\right) \text{ \ \ , \ \ }\chi \left( \overrightarrow{x}\right) =\left( 
\begin{array}{c}
\chi _{+} \\ 
\chi _{-}
\end{array}
\right)  \label{8}
\end{equation}
Eq.$\left( 7\right) $ takes the following form 
\begin{equation}
\left( 
\begin{array}{cccc}
(E-m) & 0 & -P_{z} & -\pi _{-} \\ 
0 & (E-m) & -\pi _{+} & P_{z} \\ 
-P_{z} & -\pi _{-} & (E+m) & 0 \\ 
-\pi _{+} & P_{z} & 0 & (E+m)
\end{array}
\right) \left( 
\begin{array}{c}
\varphi _{+} \\ 
\varphi _{-} \\ 
\chi _{+} \\ 
\chi _{-}
\end{array}
\right) =0  \label{9}
\end{equation}
where $\pm $ indices represent the positive and negative frequency and $\pi
_{+}$ and $\pi _{-}$ operators are defined as 
\begin{eqnarray}
\pi _{+} &=&\pi _{1}+i\pi _{2}  \label{10} \\
\pi _{-} &=&\pi _{1}-i\pi _{2}\text{ \ \ .}  \label{11}
\end{eqnarray}
The commutation relation between these operators is given as follow 
\begin{equation}
\left[ \frac{\pi _{+}}{\sqrt{2eB}},\frac{\pi _{-}}{\sqrt{2eB}}\right] =1%
\text{ .}
\end{equation}
This is similar to the commutation relation of harmonic oscillator creation
and annihilation operators: 
\begin{equation}
\left[ a^{+},a\right] =1
\end{equation}
Because of this similarity we can use the harmonic oscillator solutions,
which are the Hermite polynomials, to obtain the energy levels. The effect
of the creation and annihilation operators of our system on Hermite
polynomials is given as 
\begin{eqnarray}
\frac{\pi _{+}}{\sqrt{2eB}} &\mid &n\rangle =\sqrt{\left( n+1\right) }\mid
n+1\rangle \\
\frac{\pi _{-}}{\sqrt{2eB}} &\mid &n\rangle =\sqrt{n}\mid n-1\rangle \text{
\ .}
\end{eqnarray}

If we choose the wave function in the form 
\begin{equation}
\Psi \left( x\right) =\left( 
\begin{array}{c}
\varphi _{+} \\ 
\varphi _{-} \\ 
\chi _{+} \\ 
\chi _{-}
\end{array}
\right) =\left( 
\begin{array}{c}
a_{1}\mid n-1\rangle \\ 
a_{0}\mid n\rangle \\ 
a_{\widetilde{0}}\mid n-1\rangle \\ 
a_{2}\mid n\rangle
\end{array}
\right)
\end{equation}
and use it in Eq.$\left( 9\right) $ we obtain four linear algebraic
equations as follows: 
\begin{eqnarray}
\left( E-m\right) a_{1} &\mid &n-1\rangle -P_{3}a_{\widetilde{0}}\mid
n-1\rangle -a_{2}\sqrt{2eBn}\mid n-1\rangle =0 \\
\left( E-m\right) a_{0} &\mid &n\rangle +P_{3}a_{2}\mid n\rangle -a_{%
\widetilde{0}}\sqrt{2eBn}\mid n\rangle =0 \\
\left( E+m\right) a_{\widetilde{0}} &\mid &n-1\rangle -P_{3}a_{1}\mid
n-1\rangle -a_{0}\sqrt{2eBn}\mid n-1\rangle =0 \\
\left( E+m\right) a_{2} &\mid &n\rangle +P_{3}a_{0}\mid n\rangle -a_{1}\sqrt{%
2eBn}\mid n\rangle =0
\end{eqnarray}
These equations can be written in the matrix form as 
\begin{equation}
\left( 
\begin{array}{cccc}
\left( E-m\right) & 0 & -P_{3} & -\sqrt{2eBn} \\ 
0 & \left( E-m\right) & -\sqrt{2eBn} & P_{3} \\ 
-P_{3} & -\sqrt{2eBn} & \left( E+m\right) & 0 \\ 
-\sqrt{2eBn} & P_{3} & 0 & \left( E+m\right)
\end{array}
\right) \left( 
\begin{array}{c}
a_{1} \\ 
a_{0} \\ 
a_{\widetilde{0}} \\ 
a_{2}
\end{array}
\right) =0
\end{equation}

Consequently the allowed values of the energy are obtained by setting the
determinant of the coefficients equal to zero: 
\begin{equation}
E=\pm \sqrt{m^{2}+P_{3}^{2}+2eBn}
\end{equation}

\bigskip

\section{ENERGY SPECTRUM OF A MASSIVE SPIN-1 PARTICLE}

For a massive spin-1 particle the Kemmer equation is given in the form 
\begin{equation}
(\beta ^{\mu }\pi _{\mu }-M)\Psi _{K}\left( x\right) =0
\end{equation}
where $16\times 16$ Kemmer matrices $\beta ^{\mu }$ are given as 
\begin{equation}
\beta ^{\mu }=\gamma ^{\mu }\otimes \text{I }+\text{I}\otimes \gamma ^{\mu }
\end{equation}
with usual $\gamma ^{\mu }$ Dirac matrices. $M$ is the total mass of two
identical spin-1/2 particles and $\Psi _{K}\left( x\right) $ is the
sixteen-component wave function of the Kemmer equation. Since massive spin-1
particle will be considered as two-identical spin-1/2 particles, the wave
function can be rewritten in the form 
\begin{equation}
\Psi _{K}\left( x\right) =\Psi _{D}\otimes \Psi _{D}=\left( 
\begin{array}{c}
A \\ 
B \\ 
C \\ 
D
\end{array}
\right)
\end{equation}
where $\Psi _{D}$ is the solution of the Dirac equation and 4-component
spinors A, B, C, D are given as 
\begin{eqnarray}
A &=&\left( 
\begin{array}{c}
A_{1} \\ 
A_{0} \\ 
A_{\widetilde{0}} \\ 
A_{2}
\end{array}
\right) =\left( 
\begin{array}{c}
\varphi _{+}\varphi _{+} \\ 
\varphi _{+}\varphi _{-} \\ 
\varphi _{-}\varphi _{+} \\ 
\varphi _{-}\varphi _{-}
\end{array}
\right) \text{ \ \ \ \ \ \ \ , \ \ \ \ \ \ \ }B=\left( 
\begin{array}{c}
B_{1} \\ 
B_{0} \\ 
B_{\widetilde{0}} \\ 
B_{2}
\end{array}
\right) =\left( 
\begin{array}{c}
\varphi _{+}\chi _{+} \\ 
\varphi _{+}\chi _{-} \\ 
\varphi _{-}\chi _{+} \\ 
\varphi _{-}\chi _{-}
\end{array}
\right)  \nonumber \\
C &=&\left( 
\begin{array}{c}
C_{1} \\ 
C_{0} \\ 
C_{\widetilde{0}} \\ 
C_{2}
\end{array}
\right) =\left( 
\begin{array}{c}
\chi _{+}\varphi _{+} \\ 
\chi _{+}\varphi _{-} \\ 
\chi _{-}\varphi _{+} \\ 
\chi _{-}\varphi _{-}
\end{array}
\right) \text{ \ \ \ \ \ \ \ , \ \ \ \ \ \ \ }D=\left( 
\begin{array}{c}
D_{1} \\ 
D_{0} \\ 
D_{\widetilde{0}} \\ 
D_{2}
\end{array}
\right) =\left( 
\begin{array}{c}
\chi _{+}\chi _{+} \\ 
\chi _{+}\chi _{-} \\ 
\chi _{-}\chi _{+} \\ 
\chi _{-}\chi _{-}
\end{array}
\right) \text{.}
\end{eqnarray}
After some algebra Kemmer equation takes the following form: 
\begin{equation}
\lbrack (\gamma _{1}^{0}\otimes \text{I}+\text{I}\otimes \gamma _{2}^{0})\pi
_{0}-(\overrightarrow{\alpha }_{1}\otimes \gamma _{2}^{0}+\gamma
_{2}^{0}\otimes \overrightarrow{\alpha }_{2})\cdot \overrightarrow{\pi }%
-M\gamma _{1}^{0}\otimes \gamma _{2}^{0}]\left( 
\begin{array}{c}
A \\ 
B \\ 
C \\ 
D
\end{array}
\right) =0.
\end{equation}
We obtain four linear algebraic equations, ultimately sixteen equations from
Eq.$\left( 27\right) $: 
\begin{eqnarray}
\left( 2E-M\right) A-\overrightarrow{\sigma }_{\left( 2\right) }\cdot 
\overrightarrow{\pi }B-\overrightarrow{\sigma }_{\left( 1\right) }\cdot 
\overrightarrow{\pi }C &=&0 \\
\overrightarrow{\sigma }_{\left( 2\right) }\cdot \overrightarrow{\pi }A-MB-%
\overrightarrow{\sigma }_{\left( 1\right) }\cdot \overrightarrow{\pi }D &=&0
\\
\overrightarrow{\sigma }_{\left( 1\right) }\cdot \overrightarrow{\pi }A-MC-%
\overrightarrow{\sigma }_{\left( 2\right) }\cdot \overrightarrow{\pi }D &=&0
\\
\left( 2E+M\right) D-\overrightarrow{\sigma }_{\left( 1\right) }\cdot 
\overrightarrow{\pi }B-\overrightarrow{\sigma }_{\left( 2\right) }\cdot 
\overrightarrow{\pi }C &=&0
\end{eqnarray}
where 
\begin{equation}
\overrightarrow{\sigma }_{\left( 1\right) }\cdot \overrightarrow{\pi }%
=\left( \overrightarrow{\sigma }\otimes \text{I}\right) \cdot 
\overrightarrow{\pi }=\left[ 
\begin{array}{cccc}
P_{z} & 0 & \pi _{-} & 0 \\ 
0 & P_{z} & 0 & \pi _{-} \\ 
\pi _{+} & 0 & -P_{z} & 0 \\ 
0 & \pi _{+} & 0 & -P_{z}
\end{array}
\right]
\end{equation}
and 
\begin{equation}
\overrightarrow{\sigma }_{\left( 2\right) }\cdot \overrightarrow{\pi }%
=\left( \text{I }\otimes \overrightarrow{\sigma }\right) \cdot 
\overrightarrow{\pi }=\left[ 
\begin{array}{cccc}
P_{z} & \pi _{-} & 0 & 0 \\ 
\pi _{+} & -P_{z} & 0 & 0 \\ 
0 & 0 & P_{z} & \pi _{-} \\ 
0 & 0 & \pi _{+} & -P_{z}
\end{array}
\right] .
\end{equation}
By considering the 
\begin{eqnarray*}
A_{0} &=&A_{\widetilde{0}}\text{ \ \ , \ \ }B_{0}=C_{\widetilde{0}} \\
B_{\widetilde{0}} &=&C_{0}\text{ \ \ , \ \ }D_{0}=D_{\widetilde{0}}
\end{eqnarray*}
equalities, we can choose the A, B, C, D spinors in terms of harmonic
oscillator solutions as follows: 
\begin{eqnarray}
A &=&\left( 
\begin{array}{c}
a_{1}\mid n\rangle \\ 
a_{0}\mid n+1\rangle \\ 
a_{0}\mid n+1\rangle \\ 
a_{2}\mid n+2\rangle
\end{array}
\right) \text{ \ \ \ \ , \ \ \ \ }B=\left( 
\begin{array}{c}
b_{1}\mid n\rangle \\ 
\mid n+1\rangle \\ 
-\mid n+1\rangle \\ 
b_{2}\mid n+2\rangle
\end{array}
\right)  \nonumber \\
C &=&\left( 
\begin{array}{c}
b_{1}\mid n\rangle \\ 
-\mid n+1\rangle \\ 
\mid n+1\rangle \\ 
b_{2}\mid n+2\rangle
\end{array}
\right) \text{ \ \ \ \ , \ \ \ \ }D=\left( 
\begin{array}{c}
d_{1}\mid n\rangle \\ 
d_{0}\mid n+1\rangle \\ 
d_{0}\mid n+1\rangle \\ 
d_{2}\mid n+2\rangle
\end{array}
\right) .
\end{eqnarray}

If we use these solutions in Kemmer equation we obtain sixteen equations.
Since we have considered the spin-1 particle as a system of two-spin-1/2
particles we must eliminate the spin-0 cases due to the spin orientations.
The same equations of these sixteen equations are interpreted as
corresponding to spin-0 case, so we obtain ten equations as follows for
spin-1 particle by taking only one of \ the same equations: 
\begin{eqnarray}
\left( 2E-M\right) a_{1} &\mid &n\rangle -2P_{z}b_{1}\mid n\rangle -2\pi
_{-}\mid n+1\rangle =0 \\
\left( 2E-M\right) a_{0} &\mid &n+1\rangle -\pi _{+}b_{1}\mid n\rangle -\pi
_{-}b_{2}\mid n+2\rangle +2P_{z}\mid n+1\rangle =0 \\
\left( 2E-M\right) a_{2} &\mid &n+2\rangle +\pi _{+}\mid n+1\rangle
+P_{z}b_{2}\mid n+2\rangle =0
\end{eqnarray}
\begin{eqnarray}
P_{z}a_{1} &\mid &n\rangle -P_{z}d_{1}\mid n\rangle +\pi _{-}a_{0}\mid
n+1\rangle -\pi _{-}d_{0}\mid n+1\rangle -Mb_{1}\mid n\rangle =0 \\
\pi _{+}a_{1} &\mid &n\rangle -\pi _{-}d_{2}\mid n+2\rangle -P_{z}a_{0}\mid
n+1\rangle -P_{z}d_{0}\mid n+1\rangle -M\mid n+1\rangle =0 \\
P_{z}a_{0} &\mid &n+1\rangle +P_{z}d_{0}\mid n+1\rangle +\pi _{-}a_{2}\mid
n+2\rangle -\pi _{+}d_{1}\mid n\rangle +M\mid n+1\rangle =0 \\
\pi _{+}a_{0} &\mid &n+1\rangle -\pi _{+}d_{0}\mid n+1\rangle
+P_{z}d_{2}\mid n+2\rangle -P_{z}a_{2}\mid n+2\rangle -Mb_{2}\mid n+2\rangle
=0
\end{eqnarray}
\begin{eqnarray}
\left( 2E+M\right) d_{1} &\mid &n\rangle -2P_{z}b_{1}\mid n\rangle +2\pi
_{-}\mid n+1\rangle =0 \\
\left( 2E+M\right) d_{0} &\mid &n+1\rangle -\pi _{+}b_{1}\mid n\rangle -\pi
_{-}b_{2}\mid n+2\rangle -2P_{z}\mid n+1\rangle =0 \\
\left( 2E+M\right) d_{2} &\mid &n+2\rangle -\pi _{+}\mid n+1\rangle
+P_{z}b_{2}\mid n+2\rangle =0
\end{eqnarray}

From the first and last three equations we obtain\bigskip 
\begin{eqnarray}
a_{1} &\mid &n\rangle =\frac{1}{E-\frac{M}{2}}\left[ P_{z}b_{1}\mid n\rangle
+\pi _{-}\mid n+1\rangle \right] \\
a_{0} &\mid &n+1\rangle =\frac{1}{2(E-\frac{M}{2})}\left[ \pi _{+}b_{1}\mid
n\rangle +\pi _{-}b_{2}\mid n+2\rangle -2P_{z}\mid n+1\rangle \right] \\
a_{2} &\mid &n+2\rangle =\frac{1}{E-\frac{M}{2}}\left[ -\pi _{+}\mid
n+1\rangle -P_{z}b_{2}\mid n+2\rangle \right] \\
d_{1} &\mid &n\rangle =\frac{1}{E+\frac{M}{2}}\left[ P_{z}b_{1}\mid n\rangle
-\pi _{-}\mid n+1\rangle \right] \\
d_{0} &\mid &n+1\rangle =\frac{1}{2(E+\frac{M}{2})}\left[ \pi _{-}b_{2}\mid
n+2\rangle +\pi _{+}b_{1}\mid n\rangle +2P_{z}\mid n+1\rangle \right] \\
d_{2} &\mid &n+2\rangle =\frac{1}{E+\frac{M}{2}}\left[ \pi _{+}\mid
n+1\rangle -P_{z}b_{2}\mid n+2\rangle \right] \text{ \ .}
\end{eqnarray}
Using these equations in the others yields the following algebraic
equations: 
\begin{eqnarray}
\left[ E^{2}-\frac{M^{2}}{4}-P_{z}^{2}-eB\left( n+1\right) \right] b_{1}-%
\left[ eB\sqrt{\left( n+1\right) \left( n+2\right) }\right] b_{2} &=&0 \\
\left[ E^{2}-\frac{M^{2}}{4}-P_{z}^{2}-eB\left( n+2\right) \right] b_{2}-%
\left[ eB\sqrt{\left( n+1\right) \left( n+2\right) }\right] b_{1} &=&0
\end{eqnarray}
\begin{eqnarray}
\left[ \frac{P_{z}}{2}\sqrt{2eB\left( n+1\right) }\right] b_{1}-\left[ \frac{%
P_{z}}{2}\sqrt{2eB\left( n+2\right) }\right] b_{2} &=&\left[ E^{2}-\frac{%
M^{2}}{4}-P_{z}^{2}-eB\left( 2n+3\right) +\frac{2eBE}{M}\right] \\
\left[ \frac{P_{z}}{2}\sqrt{2eB\left( n+2\right) }\right] b_{2}-\left[ \frac{%
P_{z}}{2}\sqrt{2eB\left( n+1\right) }\right] b_{1} &=&\left[ E^{2}-\frac{%
M^{2}}{4}-P_{z}^{2}-eB\left( 2n+3\right) -\frac{2eBE}{M}\right]
\end{eqnarray}
From these algebraic equations we find the following energy spectra for
spin-1 particle moving in a homogeneous magnetic field: 
\begin{eqnarray}
E &=&\pm \frac{1}{2}\sqrt{M^{2}+4P_{z}^{2}+8eB\left( n+\frac{3}{2}\right) }
\label{55} \\
E &=&\pm \frac{eB}{M}\pm \frac{1}{2}\sqrt{(\frac{2eB}{M}%
)^{2}+M^{2}+4P_{z}^{2}+8eB\left( n+\frac{3}{2}\right) }  \label{56}
\end{eqnarray}
The normalization coefficients $b_{1}$ and $b_{2}$ are 
\begin{eqnarray}
b_{1} &=&\frac{eB\sqrt{\left( n+1\right) \left( n+2\right) }\left[ E^{2}-%
\frac{M^{2}}{4}-P_{z}^{2}-eB\left( 2n+3\right) +\frac{2eBE}{M}\right] }{%
\frac{P_{z}}{2}\sqrt{2eB\left( n+2\right) }\left[ 2eB\left( n+1\right)
-E^{2}+\frac{M^{2}}{4}+P_{z}^{2}\right] } \\
b_{2} &=&\frac{\left[ E^{2}-\frac{M^{2}}{4}-P_{z}^{2}-eB\left( n+1\right) %
\right] \left[ E^{2}-\frac{M^{2}}{4}-P_{z}^{2}-eB\left( 2n+3\right) +\frac{%
2eBE}{M}\right] }{\frac{P_{z}}{2}\sqrt{2eB\left( n+2\right) }\left[
2eB\left( n+1\right) -E^{2}+\frac{M^{2}}{4}+P_{z}^{2}\right] }\text{ \ \ \ .}
\end{eqnarray}
By using these coefficients in Eqs.$\left( 45\right) -\left( 50\right) $ one
can find the other normalization coefficients.

\section{CONCLUSIONS}

We used here a simple algebraic method for obtaining the energy levels of a
massive spin-1 particle moving in a homogeneous magnetic field. Spin-1
particle was considered as a two-identical fermion system and in the process
to obtain the energy levels the wave function was written in terms of the
harmonic oscillator solutions. Since the x-dependent vector potential
choosed in the y-direction the massive spin-1 particle will oscillate in the
x-direction and going forward to y-direction. So the solutions of the
harmonic oscillator can be used in the obtaining of the energy spectra.

The nonrelativistic approximation applies when the energies $P_{z}^{2}/M$
and $eB/M$ are small compared to the rest energy $M$. On taking the positive
roots of $E^{2}$ to first order only, one finds, as approximations to Eqs.$%
\left( 55-56\right) $ 
\begin{eqnarray}
E &=&M+\frac{P_{z}^{2}}{2M}+\frac{eB}{2M}\left( 2n+3\right) \\
E &=&M+\frac{P_{z}^{2}}{2M}+\frac{eB}{2M}\left( 2n+3\right) +\frac{eB}{2M}%
\left( \pm 1\right) \text{ \ .}
\end{eqnarray}
These are the exact results those obtained by Krase, Lu and Good . The first
of this energy spectra corresponds to $m_{s}=0$ and the second one
corresponds to $m_{s}=\pm 1$ . The same energy spectra have been also
obtained by writing the wave function in terms of Laguerre Polynomials \cite
{7}. Our result are more general than the results obtained by Wu-yang Tsai
and Asim Y\i ld\i z . They found the energy spectra in the limit $\pi _{3}=0$%
. The spectra we found are also in agreement with the spectra they found in $%
P_{z}=0$ case.

The $\pm \frac{eB}{M}$ term in the energy spectra is the spin-magnetic field
interaction term. This term shows that the magnetic moment of the spin-1
particle moving in a constant magnetic field is 
\begin{equation}
\mu _{s}=\left( \pm 1\right) \frac{e}{M}
\end{equation}

\begin{acknowledgement}
The authors thank to Professor N \"{U}nal for helpful discussions.
\end{acknowledgement}

\end{document}